\documentclass[dvipsnames,12pt]{article}
\usepackage{adjustbox}
\usepackage{amssymb}
\usepackage{amsthm}
\usepackage{amsmath}
\usepackage{authblk}
\usepackage{blkarray}
\usepackage{booktabs}
\usepackage{bm}
\usepackage{calc}
\usepackage{caption}
\usepackage{cases}
\usepackage{centernot}
\usepackage[utf8]{inputenc}
\usepackage{dcolumn}
\usepackage{float}
\usepackage[margin = 1in]{geometry}
\usepackage{graphicx}
\usepackage{lineno}
\usepackage{listings}
\usepackage{makecell}
\usepackage{mathtools}
\usepackage[authoryear,sort]{natbib}
\usepackage{pifont}

\usepackage{setspace}
\usepackage{soul}
\usepackage{subcaption}
\usepackage{tabularx}
\usepackage{tgpagella}
\usepackage{verbatim}
\usepackage{xcolor}
\usepackage{xfrac}
\usepackage{todonotes}
\usepackage[style=english]{csquotes}

\newcommand{\ve}{\varepsilon}
\newcommand{\dd}{\mathrm{d}} 

\title{Synchronized disease and behavioural dynamics in weakly coupled populations}
\author[1]{Xinxuan Wang}
\author[2]{Youngmin Park}
\author[1]{Bryce Morsky}
\affil[1]{Department of Mathematics, Florida State University, Tallahassee, FL, USA}
\affil[2]{Department of Mathematics, University of Florida, Gainesville, FL, USA}
\date{\today}

\begin{document}

\maketitle

\begin{abstract}
    The spread of infectious disease is strongly influenced by social dynamics. In addition to infection risk, individuals’ vaccination decisions depend on prevailing social behavior: high infection levels and widespread vaccination can increase vaccine uptake, which in turn suppresses infection. This feedback can generate sustained oscillations in disease prevalence and vaccination behavior. Here, we study two such populations undergoing the same behavioral-epidemiological limit cycle and introduce weak coupling between them through social influence. We show that coupling leads to synchronization of disease dynamics between the two groups. Moreover, we find that different payoff sensitivity may lead to synchronization or anti-synchronization.
\end{abstract}


\section{Introduction}

Human behaviour is a key driver of the spread of infectious disease \cite{funk10,bauch12,weston18}. Examples include mask wearing, adherence to social distancing recommendations, and vaccine hesitancy. These behavioural factors are often endogenous, changing with the state of the epidemic and social environment. Awareness of the state of the epidemic --- including disease prevalence and the death rate --- along with social norms are key drivers of such processes \cite{funk09}. In weighing the risks and benefits of taking precautions against disease, social norms can be determinative by enforcing behavioural conformity through social pressure. Examples of norms shaping decision-making include whether or not to mask \cite{yang22,heiman23,lipsey23} or get vaccinated \cite{rabb22,fayaz23}.  

There has been much work, particularly driven by the COVID-19 pandemic, to incorporate such factors into compartmental models of disease spread \cite{glaubitz20,weitz20,qiu22,ryan24,pant24,kaur25}. Such models show how behavioural responses to an epidemic can cause oscillations in infectious individuals, non-monotonic responses to changes in key variables, and suboptimal outcomes due to free-rider effects or delayed vaccine uptake. How human behaviour affects heterogeneous populations, however, is less understood. Risks of infections and social pressure may vary locally. Where the local risk of infection is relatively high, local vaccination uptake is high, which in turn prevents the spread of disease to other regions. This effect, known as ring vaccination, can be effective in reducing the prevalence of disease \cite{perisic09,wells15}. Spatially varying social dynamics may also have an impact on disease spread by disrupting social dynamics.

Social norms and dynamics can cause inadequate adherence to non-pharmaceutical interventions (NPIs) and mistimed vaccine uptake through social norm ``stickiness" \cite{morsky23,morsky25}. Due to social pressure to conform, individuals delay in changing their behaviour and thus adopt NPIs and get vaccinated later than they should given the state of the epidemic. This in turn can exacerbate the spread of disease. Simulation results from these previous models have shown that norm stickiness is promoted by shared information about the state of the epidemic. When individuals all have the same beliefs about the state of the epidemic and feel the same social pressure, they inadvertently coordinate their behaviour. This collective action can result in large and rapid shifts in opinion and behaviour. Yet, when information is constrained locally, outbreaks are smaller and dispersed over time and space. The cause of this discrepancy is a synchronization of behaviour. Thus, it is crucial to understand the conditions under which populations synchronize their behaviours, and study the outcomes.

To explore these questions of population synchronization and public health outcomes, we apply weak coupling theory to our compartmental behavioral-epidemiological models. Weak coupling theory reduces two weakly interacting oscillators (such as two populations undergoing recurring epidemics that interact through weak immigration and emigration) to a scalar equation representing the phase difference of the oscillators \cite{kuramoto1984chemical}. Existence and stability of synchronization can then be much more easily established by studying this equation \cite{ermentrout2010mathematical}. For instance, stable synchrony corresponds to zero being a stable state in the scalar phase difference equation.

Here we use weak coupling theory to study the disease dynamics between two subpopulation each in their own orbit of limit cycles. We consider three types of coupling: i) social only, ii) physical only, and iii) social and physical. In model with only social coupling, the subpopulations are coupled only through the pressure to conform from the social norm. We found that in-phase and anti-phase synchronization are both possible equilibria. In model with only physical coupling, the subpopulations are coupled physically so that infection may spread between them, but social pressure is only determined locally. In both this case and when coupling is both physical and social, we found that in-phase synchronization is the only stable equilibrium.

\section{Methods}

We present three different models of two coupled populations enduring an epidemic: one with only social coupling, one with physical coupling, and one with both. To begin, let $S_j$, $I_j$, and $P_j$ be the proportions of susceptible, infectious, and (temporarily) protected individuals in subpopulation $j$ with $S_j+I_j+P_j=1$. The protected population includes both individuals that have natural immunity as well those that have been vaccinated. Since immunity wanes at rate $\alpha$, this protection is temporary.

Individuals flow from susceptible to infectious and infectious to protected compartments in the standard way in SIRS models. $\beta$ and $\gamma$ are the transmission and recovery rates, respectively. We also include vaccination, the rate of which changes due to individual decision-making. $\eta$ is the rate at which a susceptible individual who wants to be vaccinated can be vaccinated. $1/\eta$ is then the expected time to book and make a vaccination appointment. While $\eta$ is assumed to be fixed, the desire to vaccinate $v_j$ in subpopulation $j$ is dynamic. Thus, the total rate at which a susceptible individual in subpopulation $j$ is vaccinated is $\eta v_j$.

Finally, we assume that changes in $v_j$ follow a Granovetter-Schelling dynamics \citep{schelling71,granovetter78,morsky21} in which $v_j$ changes to match the best response to the current state of the system. Here we will follow the models in \cite{morsky23,morsky25}. Being vaccinated has (perceived) costs and benefits, materially, medically, and socially. Individuals therefore make their decisions based on the infection levels and opinions of others. Consider first the utility of an unvaccinated susceptible individual, which we divide into two components, material and relational. The material utility is simply the negative of the risk of being infected $-\beta$ times the frequency of infectious individuals $\tilde{I}_j$, and is negative since it's detrimental. Note that we introduce $\tilde{I}_j$ because the subpopulations may be physically interacting. Individuals receive negative relational utility when deviating from the opinions of others, which is generated by a social norm of conformity. Deviating from the norm produces guilt, which has previously been incorporated into game theory \cite{battigalli07}. We assume that this disutility from guilt increases the further one deviates from the norm. Thus, the relational utility to an unvaccinated susceptible individual is $-\delta \tilde{v}_j^2$, where $\delta>0$ is a parameter that represent the weight of social pressure in the individual's overall utility and $\tilde{v}_j^2$ is the frequency of the desire to be vaccinated. Similarly as before, we use $\tilde{v}_j^2$ because the subpopulations may interact socially. In summary, the total utility of an unvaccinated susceptible individual in subpopulation $j$ is
\begin{equation}
    u_0(\tilde{I}_j,\tilde{v}_j) = -\beta\tilde{I}_j -\delta \tilde{v}_j^2.
\end{equation}

\begin{figure}[ht!]
\centering
\includegraphics[width=0.8\textwidth]{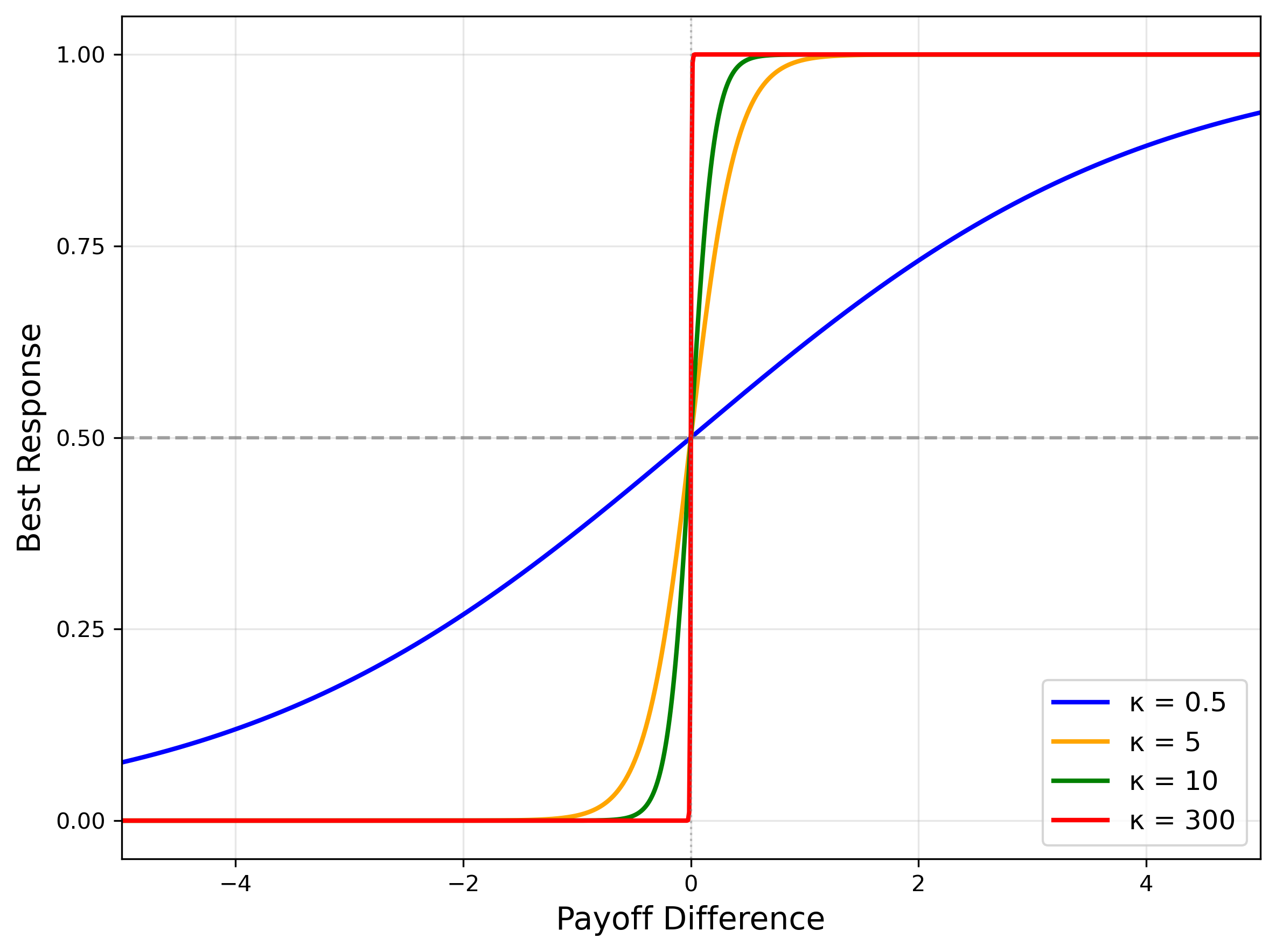}
\caption{Comparison of best response function $\sigma(\tilde{I}_j,\tilde{v}_j)$ with different values for $\kappa$.}
\label{fig:sigmoid_comparison}
\end{figure}

We assume that the vaccine, before it wanes, is completely protective, yet has a monetary cost or (perceived) medical risk $\rho>0$. The material utility is then simply $-\rho$, while the relational utility is $-\delta(1-\tilde{v}_j)^2$ for the same reasons as above. The total utility of a vaccinated individual is thus
\begin{equation}
    u_1(\tilde{I}_j,\tilde{v}_j) = -\rho -\delta(1-\tilde{v}_j)^2.
\end{equation}
Individuals compare this utility to the utility of being unvaccinated when deciding whether or not to be vaccinated with the goal of maximizing their utility. We therefore consider the utility difference
\begin{equation}
    \Delta u(\tilde{I}_j,\tilde{v}_j) = u_1(\tilde{I}_j,\tilde{v}_j) - u_0(\tilde{I}_j,\tilde{v}_j) = \beta\tilde{I}_j -\rho -\delta(1-2\tilde{v}_j).
\end{equation}
Susceptible individuals maximize their utility by being vaccinated if $\Delta u(\tilde{I}_j,\tilde{v}_j) > 0$ and remaining unvaccinated if $\Delta u(\tilde{I}_j,\tilde{v}_j) < 0$. A Heaviside function would represent the best response function for this decision-making rule. 
However, such a rule assumes perfect sensitivity to utility differences. In reality, individuals are somewhat insensitive to payoff differences due to incomplete knowledge, human errors, and reasoning ability. Therefore, we assume a smoothed best response function to model decision-making, which has the sigmoidal form
\begin{equation}
    \sigma(\tilde{I}_j,\tilde{v}_j) = \frac{1}{1 + \exp(-\kappa\Delta u(\tilde{I}_j,\tilde{v}_j))}.
\end{equation}
The parameter $\kappa$ measures the degree of sensitivity to payoff difference. A large $\kappa$ represents high sensitivity, while a small $\kappa$ represent low sensitivity. A comparison of the best response function with different $\kappa$ values is shown in Figure \ref{fig:sigmoid_comparison}.

The rate of change of $v_j$ under the Granovetter-Schelling dynamic is simply the difference between $\sigma(\tilde{I}_j,\tilde{v}_j)$ and $v_j$. Reducing dimensions due to $s_j+I_j+P_j=1$, we can finally state the complete system of differential equations for $j=1,2$:
\begin{subequations}
\begin{align}
    \dot{S}_j &= \alpha(1-S_j-I_j) - \beta S_j\tilde{I}_j- \eta v_jS_j,\\
    \dot{I}_j &= \beta S_j\tilde{I}_j- \gamma I_j,\\
    \dot{v}_j &= \sigma(\tilde{I}_j,\tilde{v}_j) - v_j.
\end{align} \label{eq:model}
\end{subequations}
Note that parameters $\alpha$, $\beta$, $\gamma$, $\rho$, and $\delta$ are assumed to not differ between subpopulations. We consider three different cases of this general model: social-only coupling, physical-only coupling, and social and physical coupling.

Under social-only coupling, individuals experience social pressure from both subpopulations, but can only be infected by members of their own subpopulation. This latter assumptions means that $\tilde{I}_j=I_j$. Regarding the former assumption, social pressure is assumed to be driven by a mixture of the $v_j$ from both population. Since coupling is weak, $\tilde{v}_j = (1-\epsilon)v_j + \epsilon v_k$ for small $\epsilon>0$ and $j \neq k$. Under physical-only coupling, individuals can be (rarely) infected by those in the other population, but only feel social pressure from their own subpopulation. With small frequency $\epsilon$, individuals physically interact with individuals in the other subpopulation, and with frequency $1-\epsilon$, they interact with members of their own subpopulation. Therefore, $\tilde{I}_j = (1-\epsilon)I_j + \epsilon I_k$ for $j \neq k$ and $\tilde{v}_j = v_j$. Note that individuals do not migrate from one subpopulation to another: subpopulations are assumed to be fixed. Rather, they only temporarily interact with the other subpopulation. Finally, when both social and physical coupling model occurs, individuals feel social pressure from both populations and can be infected by infectious from both populations. Therefore, $\tilde{I}_j = (1-\epsilon)I_j + \epsilon I_k$ and $\tilde{v}_j = (1-\epsilon)v_j + \epsilon v_k$ for small $\epsilon>0$ and $j \neq k$.

\begin{table}[!ht]
\centering
\begin{tabular}{ll}
\toprule
Parameter & Definition\\
\midrule
$\alpha = 0.01$/day & probability of loss of immunity \\
$\beta = 0.5$/day & transmission rate \\
$\gamma = 1/7$/day & probability of recovery \\
$1 \gg \epsilon > 0$ & coupling parameter \\
$\eta = 1/7$/day & rate of vaccination \\
$\delta = 0.01$ & weight of social pressure \\
$\kappa = 500$ & payoff sensitivity \\
$\rho = 0.01$ & risk of vaccine morbidity \\
\bottomrule
\end{tabular}
\caption{Parameters for numerical simulations.} \label{tbl:param}
\end{table}

We analyze the above systems using weak coupling theory and simulations, determining when subpopulations synchronize the trajectories of their epidemics and behaviour. We assume default parameters values detailed Table \ref{tbl:param}, which reflect the spread of COVID-19 \cite{bobrovitz23,byrne20,goldberg21,hassan23}. Although, our model can represent other diseases such as influenza. The time to book and make a vaccination appointment is assumed to be a week. $\kappa$, $\delta$, and $\rho$ are assumed, but in line with previous studies \cite{morsky23,morsky25}. The initial conditions for our simulations is where 1\% in subpopulation 1 is infectious and $v_1=v_2=0.01$.

\section{Results}

\subsection{Analysis of the uncoupled model}

Here we analyze the uncoupled model, building off of the results of \cite{morsky25} by conducting bifurcation analyses and analyzing the limit cycles. The uncoupled system (for $\epsilon=0$) is given by,
\begin{subequations}
\begin{align}
    \dot{S} &= \alpha(1-S-I) - \beta S I - \eta vS,\\
    \dot{I} &= \beta SI - \gamma I,\\
    \dot{v} &= \sigma(I,v) - v.
\end{align}
\end{subequations}
Note here that we dropped the subscripts. The Jacobian matrix is
\begin{equation}
    J = \left(\begin{matrix}
        -\alpha - \beta I - \eta v & -\alpha - \beta S & -\eta S \\
        \beta I & \beta S - \gamma & 0\\
        0 & \partial \sigma(I,v)/\partial I & \partial \sigma(I,v)/\partial v - 1
    \end{matrix}\right).
\end{equation}
And, $\dot{I}=0$ implies that $\bar{I} = 0$ or $\bar{S} = \gamma/\beta$. Note that $\sigma(\bar{I},\bar{v}) - \bar{v} = 0$ may have more than one solution for given $\bar{I}$ and $\bar{v}$. We thus have two scenarios, disease-free equilibria and endemic equilibria, which may have more than one equilibria.

Consider first the disease-free scenario, then $\bar{S} = \alpha/(\alpha+\bar{v})$ and the Jacobian matrix becomes
\begin{equation}
    J = \left(\begin{matrix}
        -\alpha - \eta \bar{v} & -\alpha - \beta \bar{S} & -\eta \bar{S} \\
        0 & \beta \bar{S} - \gamma & 0\\
        0 & \partial \sigma(\bar{I},\bar{v})/\partial I & \partial \sigma(\bar{I},\bar{v})/\partial v - 1
    \end{matrix}\right),
\end{equation}
which has the characteristic equation
\begin{equation}
    (\lambda + \alpha + \eta \bar{v})(\lambda - \beta \bar{S} + \gamma)(\lambda - \partial \sigma(\bar{I},\bar{v})/\partial v + 1) = 0.
\end{equation}
One eigenvalue is always negative ($\lambda = - \alpha - \eta \bar{v}$). $\lambda = \partial \sigma(\bar{I},\bar{v})/\partial v - 1$ is negative when $\bar{v}$ is large or small, but positive for intermediate values. Generally, we consider the case where $\beta>\gamma$, which gives us $R_0>0$. If $\bar{S} \approx 1$, then this condition gives us $\lambda_2 = \beta \bar{S} - \gamma \approx \beta - \gamma > 0$, and the equilibrium is unstable. However, social pressure could be sufficient to stabilize a high $\bar{v}$, and thereby a low $\bar{S}$, stabilizing such a disease-free equilibrium.

Next, consider the endemic equilibrium, where
\begin{equation}
    \bar{I} = \frac{\alpha(\beta-\gamma) - \gamma \eta \bar{v}}{\beta(\alpha+\gamma)}.
\end{equation}
The Jacobian matrix simplifies to
\begin{equation}
    J = \left(\begin{matrix}
        -\alpha - \beta\bar{I} - \eta\bar{v} & -\alpha - \gamma & -\eta \gamma/\beta \\
        \beta\bar{I} & 0 & 0\\
        0 & \partial \sigma(\bar{I},\bar{v})/\partial I & \partial \sigma(\bar{I},\bar{v})/\partial v - 1
    \end{matrix}\right),
\end{equation}
which has the characteristic equation
\begin{multline}
        \lambda^3 + \left[ 1 - \frac{\partial \sigma(\bar{I},\bar{v})}{\partial v} + \alpha + \beta\bar{I} + \eta\bar{v} \right]\lambda^2 + \left[\left( 1 - \frac{\partial \sigma(\bar{I},\bar{v})}{\partial v} \right)(\alpha + \beta\bar{I} + \eta\bar{v})+\beta(\alpha+\gamma)\bar{I}\right]\lambda \\+ \left[ \bar{I}\beta(\alpha+\gamma)\left( 1 - \frac{\partial \sigma(\bar{I},\bar{v})}{\partial v} \right) + \bar{I}\gamma\eta\frac{\partial \sigma(\bar{I},\bar{v})}{\partial I} \right] = 0.
\end{multline}
Using the parameters from Table \ref{tbl:param}, the disease-free equilibrium is unstable, since $\beta\bar{S} - \gamma > 0$. Additionally, there is an unstable endemic equilibrium and numerical simulations suggest that there is a stable limit cycle.

\begin{figure}[ht!]
    \centering
    \includegraphics[width=\linewidth]{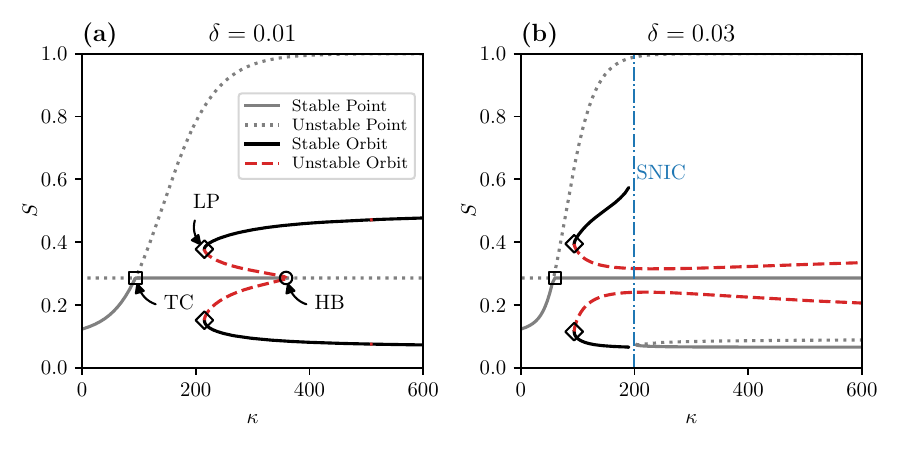}
    \caption{One-parameter bifurcation diagram in the payoff sensitivity $\kappa$ for two different values of the weight of social pressure $\delta$. (a) One-parameter bifurcation diagram for $\delta=0.01$. TC (square) denotes a transcritical bifurcation, HB (circle) denotes a (subcritical) Hopf bifurcation, and LP (diamond) denotes a limit point (fold bifurcation of cycles). (b) One-parameter bifurcation diagram for $\delta = 0.03$. The transcritical (square) and fold of cycles (diamond) remain, but the Hopf bifurcation is replaced by a saddle-node on an invariant cycle (SNIC) bifurcation (blue vertical dash dot line). The fold bifurcation of cycles (diamond) serves as a lower bound in $\kappa$ for the existence of limit cycles, and the SNIC bifurcation (vertical dash-dot line) serves as an upper bound in $\kappa$ for the existence of limit cycles.}
    \label{fig:one_parameter}
\end{figure}

To investigate the parameters for which there exists a limit cycle, we compute a pair of one-parameter bifurcation diagrams in the payoff sensitivity $\kappa$ in Figure \ref{fig:one_parameter} for two different values of the weight of social pressure $\delta=0.01,0.03$. Figure \ref{fig:one_parameter}(a) shows that for $\kappa\in[220,350]$, there exists bistability between a stable endemic equilibrium (solid gray) and a stable limit cycle solution (solid black) separated by an unstable limit cycle (dashed red). As $\kappa \rightarrow 200^+$, the stable and unstable limit cycles meet in a fold bifurcation of cycles (LP, diamond).  For $\kappa \in[100,220]$, there is one stable endemic equilibrium (solid gray) and an unstable endemic equilibrium (dotted gray). Near $\kappa \approx 100$, there is a transcritical bifurcation where the stable endemic state decreases as $\kappa$ decreases. While there is a subcritical Hopf bifurcation (HB, circle), it ultimately does not determine the interval of existence of the limit cycle. 

Figure \ref{fig:one_parameter}(b) exhibits the same system and parameters as Panel (a), but with social pressure weight increased from $\delta = 0.01$ to $\delta = 0.03$. There is no longer a Hopf bifurcation for greater $\delta$. Instead, a saddle-node on an invariant cycle (SNIC) bifurcation (blue vertical dash dot line) delimits the upper bound of existence for limit cycles in $\kappa$.

\begin{figure}[ht!]
    \centering
    \includegraphics[width=\textwidth]{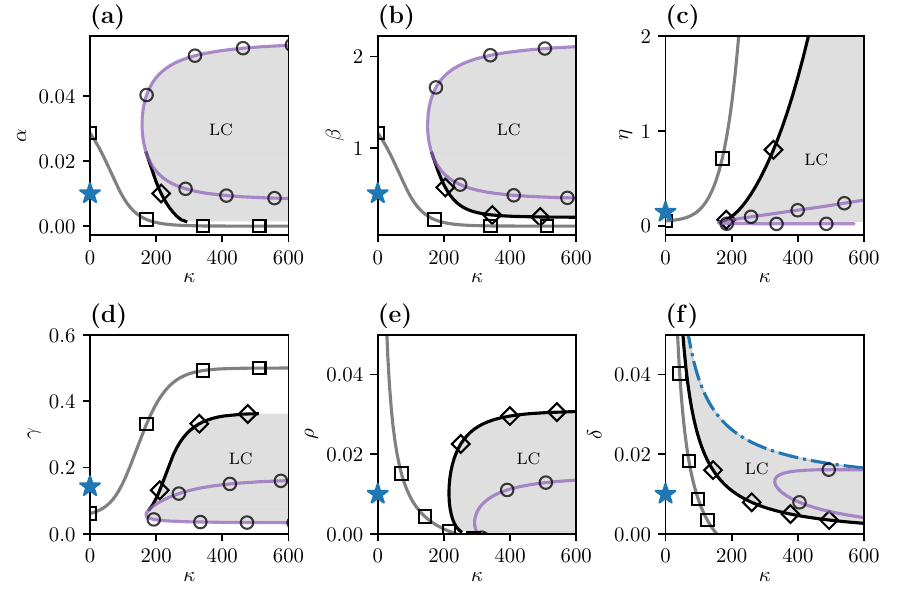}
    \caption{Two-parameter bifurcation diagram in the payoff sensitivity $\kappa$ as a function of all other model parameters ((a): $\alpha$, (b): $\beta$, (c): $\eta$, (d): $\gamma$, (e): $\rho$, and (f): $\delta$). In each panel, black curves with diamonds denote fold bifurcations of cycles (which corresponds to LP (diamond) in Figure \ref{fig:one_parameter}), purple curves with circles denote Hopf bifurcations (which corresponds to HB (circle) in Figure \ref{fig:one_parameter}), and gray curves with squares denote transcritical bifurcations (which corresponds to TC (square) in Figure \ref{fig:one_parameter}). Gray shaded regions correspond to parameter values where a stable limit cycle (LC) exists. The blue star in each panel denotes the default respective parameter value. (a)--(e): a stable limit cycle exists to the right of the fold bifurcation of cycles (black, diamonds) or the Hopf bifurcation (purple, circles). (f): A stable limit cycle exists between the fold bifurcation of cycles (black, diamonds) and the SNIC (dash dot blue).}
    \label{fig:two_parameter}
\end{figure}

By tracking the fold bifurcation of cycles (LP, diamond) in Figure \ref{fig:one_parameter} and the SNIC bifurcation (vertical blue dash-dot line) as a function of a second parameter, we may investigate the existence (or non-existence) of stable limit cycle solutions in parameter space. This process yields Figure \ref{fig:two_parameter}, where all horizontal axes correspond to the payoff sensitivity $\kappa$, while each vertical axis corresponds to a different model parameter $\alpha$, $\beta$, $\eta$, $\gamma$, $\rho$, $\delta$, respectively. These two-parameter bifurcation diagrams show that the stable limit cycle tends to exist for $\kappa$ sufficiently large, typically to the right of fold bifurcation of cycles (black, diamonds). The only exception is Figure \ref{fig:two_parameter}(f), where limit cycles terminate through a SNIC bifurcation (blue dash-dot). In Figure \ref{fig:two_parameter} Panels (a), (b), (c), and (d), there are co-dimension two points where the fold bifurcation of cycles (black, diamonds) coincide with the Hopf bifurcation (purple, circles) -- these points are called the Bautin bifurcation or generalized Hopf bifurcation \cite{kuznetsov1998elements} (Chapter 8.3).

\subsection{Phase Reduction Theory}

We briefly introduce some terminology from Phase Reduction Theory that we will use for analyzing our model. More details about everything mentioned in this section can be found at \cite{kuramoto1984chemical,ermentrout2010mathematical,Schwemmer2012,park2017utility}. We begin with an intuitive description. Suppose that the autonomous system
\begin{equation}\label{eq:dx_dt}
    \frac{\dd X}{\dd t} = F(X),
\end{equation}
has a $T$-periodic limit cycle solution $\gamma(t)$. The phase of the limit cycle oscillator  $\gamma(t)$ is given by $\theta(t) \in [0, 2\pi)$, where
\begin{equation}\label{eq:phase_constant}
    \frac{\dd \theta}{\dd t} = \omega = 2\pi/T.
\end{equation}
That is, $\theta(t)$ associates each spatial coordinate of the limit cycle to a phase variable (where the phase variable is directly proportional to time $t$). The phase variable allows us to work in angular units where the angular frequency is $\omega = 2\pi/T$ \eqref{eq:phase_constant}. In our system, we choose $\theta = 0$ to correspond to the point of peak infection. Let $\theta_1, \theta_2$ denote the phases of subpopulations $1$ and $2$, and $\phi = \theta_2 - \theta_1$ be the phase difference. We wish to study the fixed points and  stability of the phase difference dynamics $\dot{\phi}$.

The Infinitesimal Phase Response Curve (iPRC) $Z(t)= \nabla \theta(\gamma(t))$ is the gradient of the phase \cite{izhikevich2007dynamical,ermentrout2010mathematical}. It measures the phase shift due to an infinitesimally small perturbation to the limit cycle at any given phase. The iPRC satisfies the adjoint equation 
\begin{align}
    \dot{Z} = -A(t)^T \cdot Z(t), \label{adjoint}
\end{align}
where  $A(t) = DF(\gamma(t))$ \cite{malkin1949methods}. A proof that the iPRC satisfies Equation \eqref{adjoint} can be found in \cite{ermentrout2010mathematical,Schwemmer2012}. Since we are working with angular units where one full cycle is $2\pi$, we normalize $Z(t)$ so that
\begin{align}\label{eq:normalize}
    \frac{d\theta}{dt} = \nabla_X\theta\big(\gamma(t)\big) \cdot \gamma'(t) = Z(t) \cdot F\big(\gamma(t)\big) = \omega = \frac{2\pi}{T}.
\end{align}
The adjoint system \eqref{adjoint} has the opposite stability of the original system \eqref{eq:dx_dt}. Therefore, to solve for $Z(t)$, we integrate backwards in time from an arbitrary initial condition and normalize the solution using \eqref{eq:normalize} \cite{ermentrout1996type,ermentrout2003simulating}. 

\begin{figure}[ht!]
    \centering
    \includegraphics[width=\textwidth]{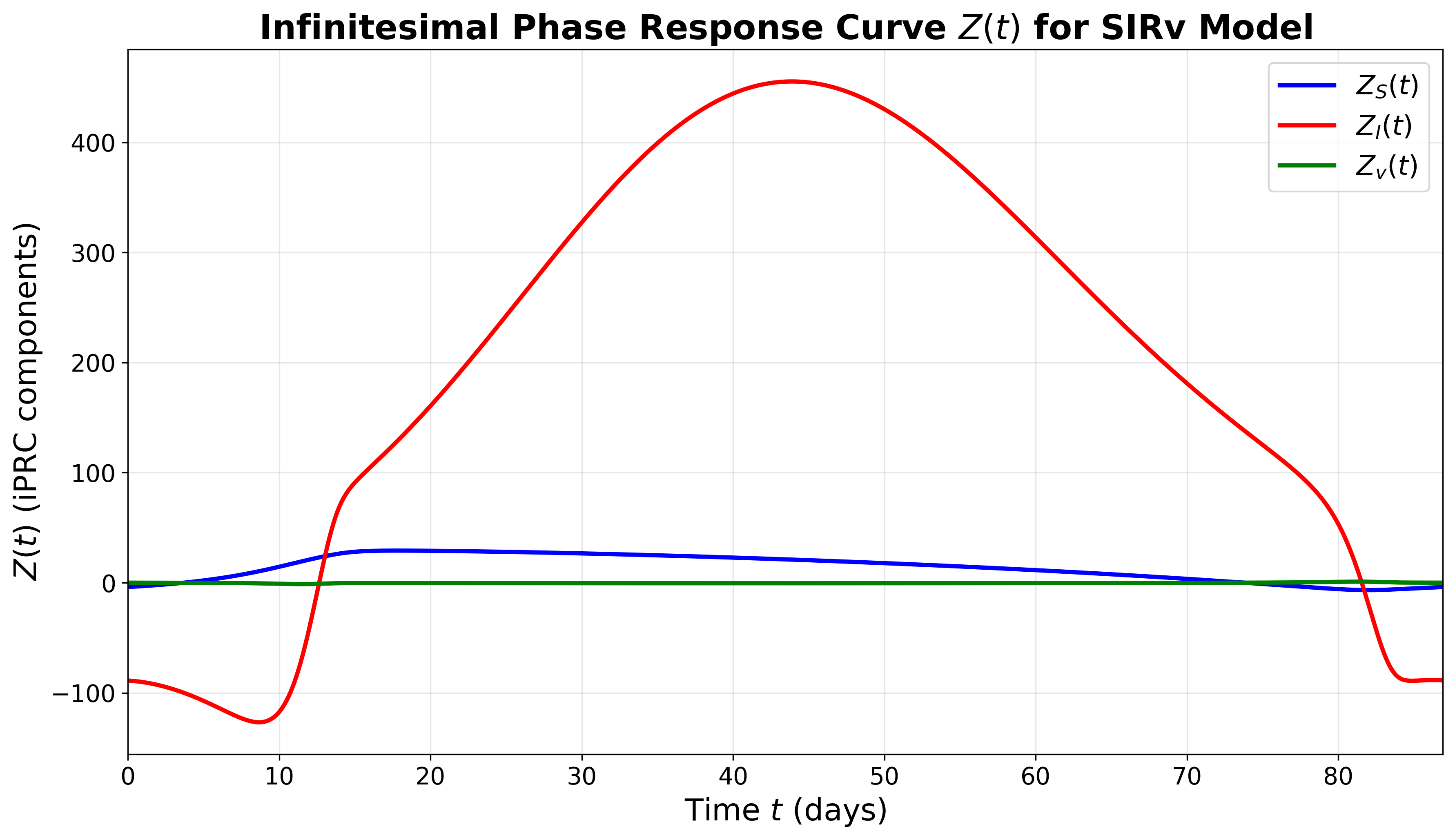}
    \caption{Components of the iPRC $Z(t) = (Z_S, Z_I, Z_v)$ for the uncoupled system, which measures the instantaneous phase change of each component by a small perturbation at time $t$.}
    \label{fig:iPRC}
\end{figure}
\begin{figure}[ht!]
    \centering
    \includegraphics[width=\textwidth]{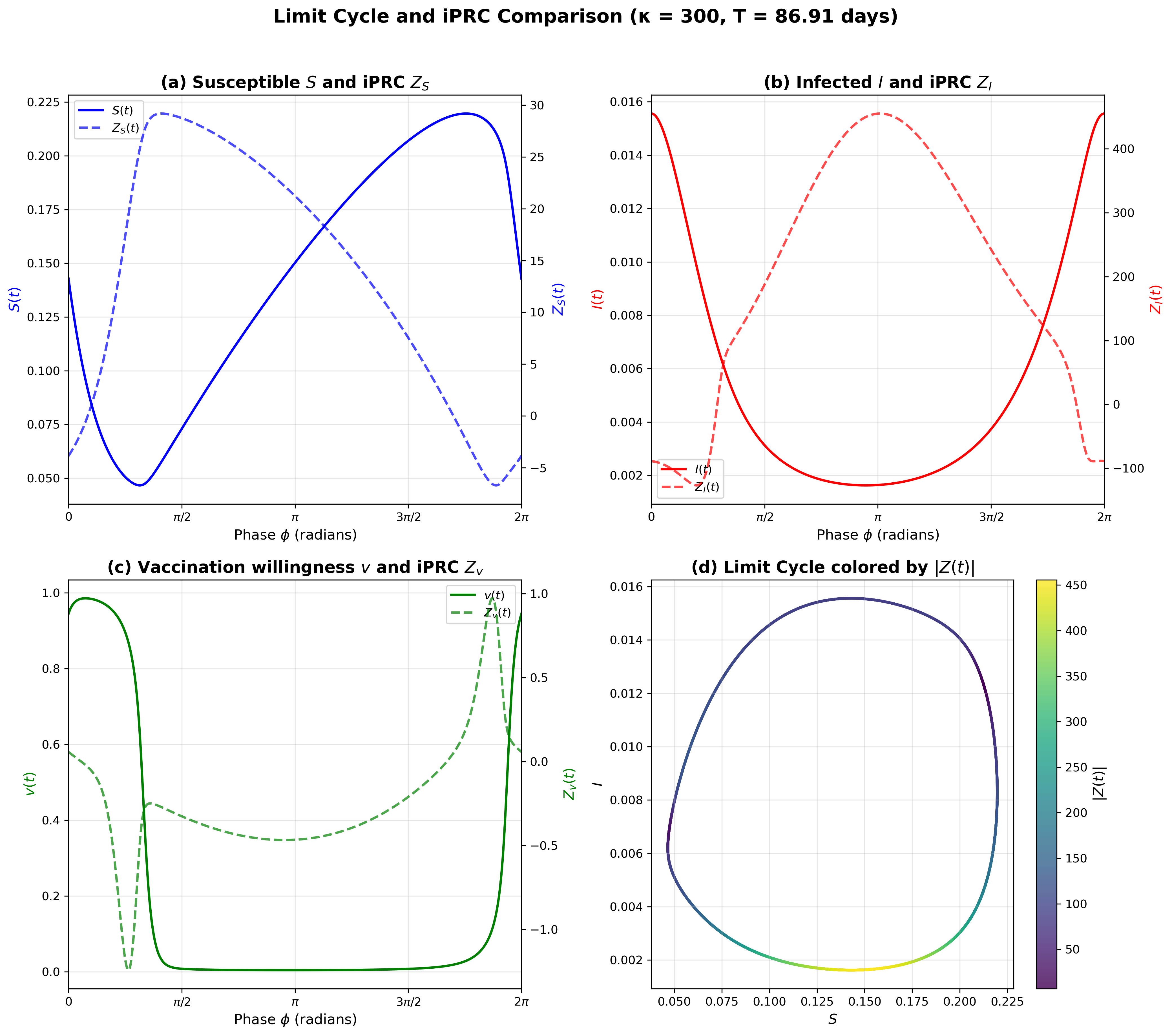}
    \caption{Limit cycle and iPRC comparison.}
    \label{fig:iPRCvsLC}
\end{figure}

By comparing the iPRC and the limit cycle in Figure \ref{fig:iPRCvsLC}, we can quantify how an infinitesimal perturbation to each state variable affects the phase of the disease dynamics. Specifically, a perturbation $\lambda x_i$ applied at phase $\theta$ to the $i$-th component $x_i = S, I$ or $v$ induces a phase shift $\Delta\theta \approx Z_i(\theta)\,\lambda x_i$, where $Z_i$ is the $i$-th component of the iPRC. These results have implications for public health interventions. For instance, in Figure \ref{fig:iPRCvsLC}(c), $Z_v(t)$ attains its minimum value of $-1.24$ at $\theta = 0.26\pi$. A positive perturbation $\lambda v > 0$ at this phase --- corresponding to a targeted campaign encouraging vaccination --- would induce a negative phase shift $\Delta\theta < 0$, effectively resetting the system to an earlier phase where vaccination willingness was higher.

Now that we have an understanding of how oscillator phases are derived, we turn to reducing the coupled system, which we express as
\begin{align}
    \frac{dX_j}{dt} = F(X_j) + \epsilon G_j(X_j, X_k), \quad j=1,2,\quad k={3-j}.\label{FG_formula}
\end{align}
We provide an abridged self-contained derivation; a detailed derivation can be found in \cite{kuramoto1984chemical,ermentrout2010mathematical,Schwemmer2012}.In the coupled case, the phase description of the oscillators can be written
\begin{equation*}
    \frac{d\theta_j}{dt} = \omega + \ve Z(\theta_j)\cdot G_j(\gamma(\theta_j),\gamma(\theta_k)).
\end{equation*}
Note that because coupling is weak, we assume that phase perturbations may be captured as phase shifts in the limit cycle $\gamma$ of the uncoupled system. This assumption is valid so long as a weakly perturbed trajectory returns to the limit cycle amplitude within one period. To simplify the calculations, we subtract the moving frame $\omega t$ from $\theta_j$ so that we are left with a pair of non-autonomous phase equations:
\begin{equation}\label{eq:non-autonomous}
    \frac{d\theta_j}{dt} = \ve Z(\theta_j+\omega t)\cdot G_j(\gamma(\theta_j+\omega t),\gamma(\theta_k+\omega t)),
\end{equation}
where we have abused notation and let $\theta_j \equiv \theta_j - \omega t$. Since we are only concerned with long-term behavior of phase differences, we average \eqref{eq:non-autonomous} over one period \cite{sanders2007averaging}, which has the added benefit of yielding an autonomous ODE:
\begin{equation}
\begin{split}
    \frac{d\theta_j}{dt} &= \frac{\ve}{T}\int_0^T Z(\theta_j+\omega t)\cdot G_j(\gamma(\theta_j+\omega t),\gamma(\theta_k+\omega t)) \,\mathrm{d}t\\
    &= \frac{\ve}{\omega T}\int_0^T Z(s)\cdot G_j(\gamma(s),\gamma(\theta_k-\theta_j+s)) \,\mathrm{d}s,
\end{split}
\end{equation}
where the second line follows from a straightforward change in variables $s = \theta_j + \omega t$. We thus obtain the phase equations,
\begin{equation}
    \omega\frac{d\theta_j}{dt} = \ve H(\theta_k-\theta_j), \quad j=1,2,\quad k=3-j,
\end{equation}
where
\begin{align}\label{eq:h0}
    H(\phi) = \frac{1}{T} \int_0^T Z(t) \cdot G\big(\gamma(t), \gamma(t+\phi)\big) dt.
\end{align}
The function \eqref{eq:h0} describes how they interact through the coupling function $G_j$ to alter oscillator phases. Finally, defining $\phi = \theta_2 - \theta_1$, we arrive at the phase difference equation,
\begin{equation}
    \dot \phi = \ve[H(\phi) - H(-\phi)] \equiv \ve C(\phi).
\end{equation}
Note that we have chosen to absorb the parameter $\omega$ into $C$ because it only scales the magnitude of the dynamics $\dot\phi$ and does not affect more important quantities like bifurcation points and the existence and stability of fixed points.




By the symmetry of $C(\phi)$ and periodicity of $H(\phi)$, we know all even derivatives of $C(\phi)$ are $0$ for any $\phi$, and that 
\begin{align}
    C(0) &= H(0) - H(0) = 0, \\
    C(\pi) &= H(\pi) - H(-\pi) = 0. \label{Cpi}
\end{align}
Therefore synchrony and antiphase are always fixed points.

To perform phase reduction analysis for our social only coupled system, recall the coupled equations
\begin{equation}\tag{\ref{eq:model}}
\begin{split}
    \dot{S}_j &= \alpha(1-S_j-I_j) - \beta S_j\tilde{I}_j- \eta v_jS_j,\\
    \dot{I}_j &= \beta S_j\tilde{I}_j- \gamma I_j,\\
    \dot{v}_j &= \sigma(\tilde{I}_j,\tilde{v}_j) - v_j,
\end{split}
\end{equation}
where
\begin{equation*}
(\tilde{I}_j,\tilde{v}_j) = \begin{cases}
    (I_j, (1-\epsilon)v_j + \epsilon v_k) & \text{(Social-only)},\\
    ((1-\epsilon)I_j + \epsilon I_k,v_j) & \text{(Physical-only)},
\end{cases}
\end{equation*}
for small $\epsilon>0$ and $j \neq k$, and
\begin{equation*}
    \sigma(\tilde{I}_j,\tilde{v}_j) = \frac{1}{1 + \exp(-\kappa\Delta u(\tilde{I}_j,\tilde{v}_j))}, \quad  \Delta u(\tilde{I}_j,\tilde{v}_j) = \beta\tilde{I}_j -\rho -\delta(1-2\tilde{v}_j).
\end{equation*}
We first need to express \eqref{eq:model} in the form of Equation (\ref{FG_formula}). To do that, consider the social-only case for concreteness. We need to expand the sigmoid function $\sigma(x) = 1/(1 + \exp(-\kappa x))$ in the $v_j$ component in a Taylor series. Let $\xi_0 = -\kappa[\beta I_j - \rho - \delta(1-2v_j)]$ and $\xi_c =-\kappa[\beta I_j - \rho - \delta (1-2((1-\epsilon)v_j+\epsilon v_k))]$, then $\dot v = \sigma(\xi_0) - v$ and $\dot{v_j} = \sigma(\xi_c) - v_j$. Taylor expanding the sigmoid function around $\xi_0$ gives us
\begin{equation}
    \sigma(\xi_c) = \sigma(\xi_0) + \sigma'(\xi_0) \cdot (\xi_c - \xi_0) + O\big((\xi_c - \xi_0)^2\big) = \sigma(\xi_0) + \sigma'(\xi_0) \epsilon \zeta + O\big((\epsilon \zeta)^2\big),
\end{equation}
where $\epsilon\zeta = \xi_c - \xi_0 = -2\epsilon \kappa \delta (v_j-v_k)$. Therefore, we have
\begin{align}
    \dot{X}_j = F(X_j) + \epsilon G(X_j, X_k), \quad j = 1,2, \quad k = 3-j,
\end{align}
where $X_j = (S_j, I_j, v_j)^T$,
\begin{equation}
    F(X_j) = \begin{pmatrix}
        \alpha(1 - S_j - I_j) - \beta S_j I_j - \eta v_j S_j \\[8pt]
        \beta S_j I_j-\gamma I_j \\[8pt]
        \sigma(\xi_0)-v_j
    \end{pmatrix},
\end{equation}
and
\begin{equation}
    G(X_j, X_k) = \begin{pmatrix}
        0 \\[8pt]
        0 \\[8pt]
        \sigma'(\xi_0)\zeta
    \end{pmatrix}.
\end{equation}
Similar calculation shows that the $G(X_j, X_k)$ function for the models with physical coupling is
\begin{equation}
    G(X_j, X_k) = \begin{pmatrix}
        \beta S_j(I_j-I_k) \\[8pt]
        S_j(I_j-I_k) \\[8pt]
        \sigma'(\xi_0)\zeta_p \label{G_j_for_other_two}
    \end{pmatrix},
\end{equation}
where $\zeta_p = \beta\kappa(I_j - I_k)$ for physical only coupling and $\zeta_p = \kappa[\beta(I_j - I_k)-2\delta(v_j-v_k)]$ for both physical and social coupling, respectively.

\begin{figure}[ht!]
    \centering
    \includegraphics[width=0.8\textwidth]{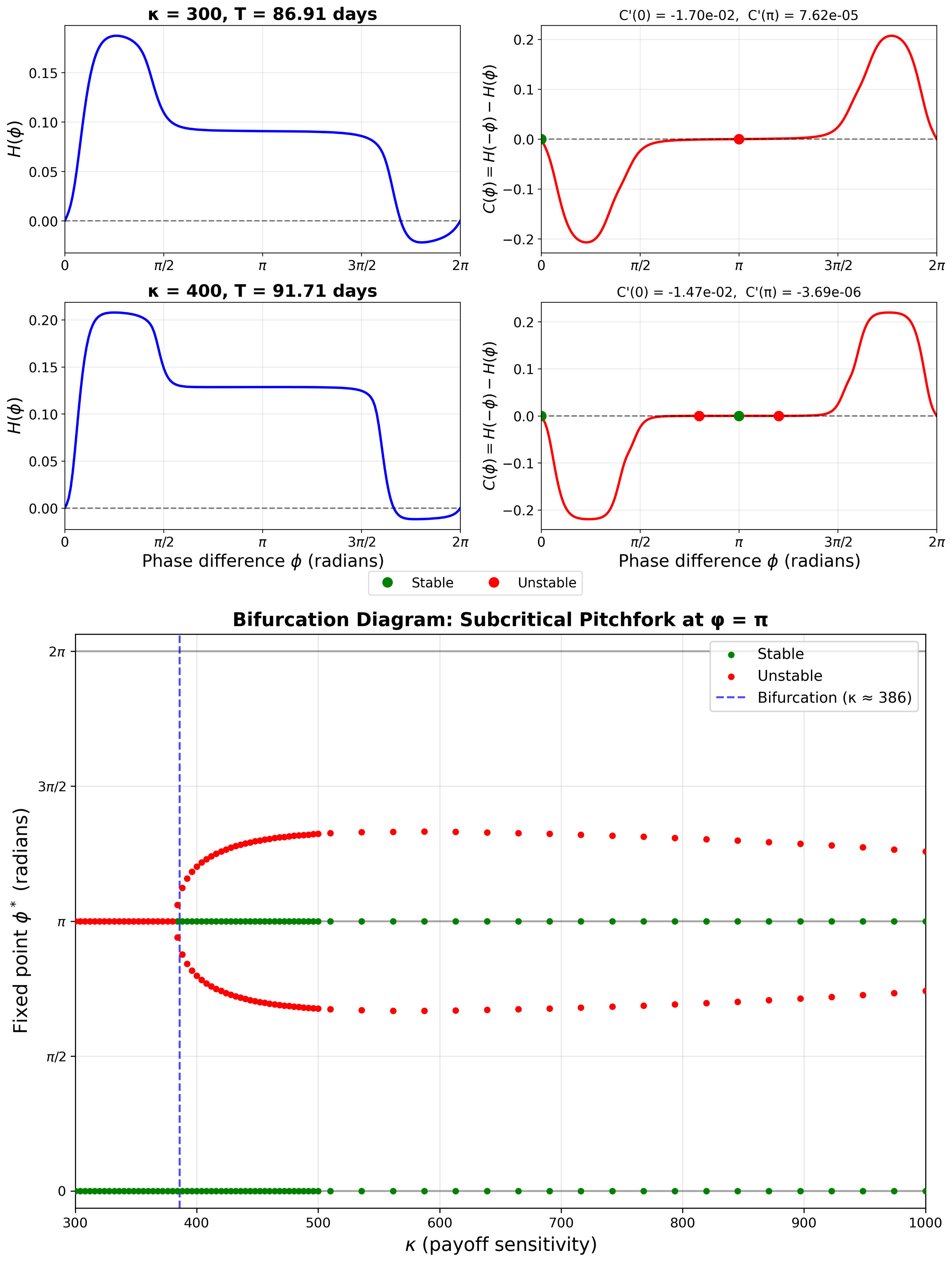}
    \caption{Coupling functions $H(\phi)$ and phase dynamics $C(\phi)$ for various values of $\kappa$ and its bifurcation diagram.}\label{fig:c_phi}
\end{figure}

Figure \ref{fig:c_phi} shows the numerical calculation of $H(\phi)$ and thus $C(\phi) = \dot{\phi}/\epsilon$ for $\kappa = 300$ and $400$. From these results we can see that $\phi = 0$, i.e. synchronization is stable across different values of $\kappa$. Further, there is a subcritical pitchfork bifurcation for $\phi = \pi $ for $\kappa$ at $\kappa^* = 386.5$. In the Appendix, we show the same numerical calculations for physical coupling and both coupling. 

\subsection{Bifurcation Analysis for $\kappa$} \label{bifurcation_kappa}

$\kappa$ is the parameter representing payoff sensitivity. The larger $\kappa$ is, the more sensitive individuals are to small payoff difference. As such, they will adjust their decisions more quickly. In effect, the sigmoid approaches a Heaviside function. A comparison of how different $\kappa$ values shape the sigmoid is shown in Figure \ref{fig:sigmoid_comparison}. We analyze the bifurcation by Taylor expansion around $\phi = \pi$ and $\kappa = \kappa^* = 386.5$, and we calculate the derivatives using either finite difference or spline.

At $\kappa^* = 386.5, C'(\pi, \kappa^*) = 0$. Let $\psi = \phi - \pi$, then for $\kappa = \kappa^*$ we have
\begin{equation}
    C(\pi+\psi) = C(\pi) + C'(\pi) \psi + \frac{C''(\pi)}{2}\psi^2 + \frac{C'''(\pi)}{6}\psi^3 + O(\psi^4) =\frac{C'''(\pi)}{6}\psi^3 + O(\psi^5),
\end{equation}
given Equation (\ref{Cpi}) and the fact that even derivatives of $C(\phi) = 0$ for any $\phi$. Numerically calculated derivatives show that $C'''(\pi)/6 = 6 \cdot 10^{-8} > 0$. Thus, at the bifurcation point $\kappa = \kappa^*$, $\psi$ is unstable. Let $k = \kappa - \kappa^*$, and expand $C'(\pi)$ around $\kappa^*$ to see how $C'(\pi)$ changes as we vary $\kappa$:
\begin{equation}
    C'(\pi, \kappa) =  C'(\pi, \kappa^*) + \frac{\partial C'(\pi, \kappa^*)}{\partial \kappa} \cdot k + O(k^2) = ak + O(k^2) ,\label{C'pik}
\end{equation}
where $a = \frac{\partial C'(\pi, \kappa^*)}{\partial \kappa}$. Expand $C(\pi+ \psi, \kappa)$ around $C(\pi, \kappa)$:
\begin{align}
    C(\pi + \psi, \kappa) = C(\pi,\kappa) + C'(\pi,\kappa) \cdot \psi + \frac{C''(\pi, \kappa)}{2} \cdot \psi^2 + \frac{C'''(\pi, \kappa)}{6} \cdot \psi^3 + O(\psi^5). \label{cphikappa}
\end{align}
Since $C(\pi,\kappa) = 0$ and $ C'(\pi, \kappa)$ was previously calculated, the only term not accounted for is $C'''(\pi, \kappa)/6$. If we expand $C'''(\pi, \kappa)$ around $\kappa^*$, we see that
\begin{align}
    C'''(\pi, \kappa) = C'''(\pi, \kappa^*) + O(k), \label{assumed}
\end{align}
which means for $k$ small, $C'''(\pi, \kappa) \approx C'''(\pi, \kappa^*)$. Let $b = C'''(\pi, \kappa^*)/6$. Then, combining $b$ with Equation (\ref{cphikappa}) and the fact that $\psi = \phi - \pi$, we have 
\begin{equation}
    C(\pi + \psi, \kappa) = \frac{1}{\epsilon}\dot \phi = \frac{1}{\epsilon} \dot \psi = ak\psi + b\psi^3.
\end{equation}
We can study the dynamics of
\begin{align}
    \frac{1}{\epsilon} \dot \psi = ak\psi + b\psi^3
\end{align}
by using spline, giving us $b = 10^{-8} > 0$. Since $a <0$, we know that there is a subcritical pitchfork bifurcation at $\kappa^* = 386.5$.

At $\kappa = \kappa^*$, i.e.\ $k = 0$, $\psi = \epsilon  b \cdot\psi^3, \ddot{\psi} = 3\epsilon \cdot \psi^2 > 0$ for any $\psi > 0$, so $\pi$ is unstable at $\kappa^*$, which we have discussed earlier in this section. For $\kappa \ne \kappa^*$, i.e.\ $k \ne 0$, we have
\begin{equation}
    \frac{1}{\epsilon} \dot \psi = ak \cdot \psi + b \cdot \psi^3 =
    \psi (ak + b \cdot \psi^2) = 0.
\end{equation}
Therefore, $\psi_0 = 0$, i.e. $\phi_0 = \pi$ is always a fixed point. For the other fixed point
\begin{equation*}
    ak + b \cdot \psi^2 = 0 \implies \psi^2 = -\frac{ak}{b}.
\end{equation*}
Since $a < 0$, $-ak/b > 0$ if $k > 0$, i.e. $\kappa > \kappa^*$, and the new fixed points will be 
\begin{equation*}
    \psi_1 = \pm \sqrt{-\frac{ak}{b}}, \quad \psi_2 = -\sqrt{-\frac{ak}{b}}.
\end{equation*}
To test their stability, note that
\begin{align*}
    \ddot \psi &= ak + 3b\cdot \psi^2\\
    \ddot \psi\bigg(\pm \sqrt{-\frac{ak}{b}}\bigg) &= ak - 3b\cdot \frac{ak}{b}\\
    &= ak-3ak\\
    &= -2ak >0 \text{ when }k>0.
\end{align*}
Therefore $\psi_1$ and $\psi_2$ are both unstable. Note that the locations of the new unstable fixed points are only accurate for $\kappa$ values near $\kappa^*$, since at equation (\ref{assumed}) we assumed $k$ to be small. We show the bifurcation diagrams for other parameters in the Appendix. 
\section{Discussion}

Here we have uncovered conditions under which the trajectories of epidemics synchronize between subpopulations, either in phase or antiphase, even when coupling is weak and there is no between-population flow of infectious individuals. We find that antiphase synchronization is stable when behavioral change is highly sensitive to payoff differences, but unstable when it is not. However, even when antiphase is stable, it may take more than $10,000$ days for the populations to synchronize to antiphase. Therefore, we suspect that antiphase synchronization is rarely observed in reality. In contrast, in-phase synchronization is always stable given that the uncoupled systems form a limit cycle. This synchronization occurs most rapidly for an intermediate degree of payoff sensitivity. 

Our results have important implications to public policy. For example, in phase synchronization magnifies total peak prevalence of infections. This effect could overwhelm hospitals and governmental responses. Additionally, selectively containing an outbreak in one subpopulation could be ineffective \cite{budich21}. On the other hand, in phase synchronization may increase the risk of extinction of the virus. When disease prevalence crashes, stochastic effects could lead to its local extinction. When local epidemics are not synchronized, neighbouring regions can serve as reservoirs of disease, allowing for reinfection. Regional synchronization, on the other hand, could lead to regional extinction. Synchronization driven by social coupling could also provide a potential lever for public policy. The timing of promoting vaccination regionally can be more precise \cite{morsky25}.

Future research could consider the effects of different social norms or models of social learning and contagion. Further, there can exist salient differences between populations such as population size, local transmission rates, and local social norms. Rural and urban population often differ intrinsically with respect to these factors, and thus is an important motivation for considering such variations. These and our studies would be strengthened by validating findings with data and further exploring their implications to public policy by designing optimal policies in light of these results. Finally, previous research has explored synchronization in epidemic \cite{lloyd96,zheng17,eclerova25} and, more generally, spatial ecological interactions \cite{ranta95,fox11,goldwyn11}. Outside of social dynamics, these phenomena can be generated by processes of dispersal \cite{goldwyn08}, noise \cite{greenman01}, seasonality, or other synchronous external factors. Addressing such processes ecology and evolution with weak coupling theory is another direction for future work.

\subsection*{Code and Data Availability}
Code to run the simulations is available at github.com/xxwang16/NID.

\bibliographystyle{plain}
\bibliography{bib/refs}

\appendix

\section{Phase Reduction for Physical Coupling and Combined Coupling}
\begin{figure}[htbp]
    \centering
    \includegraphics[width=\textwidth]{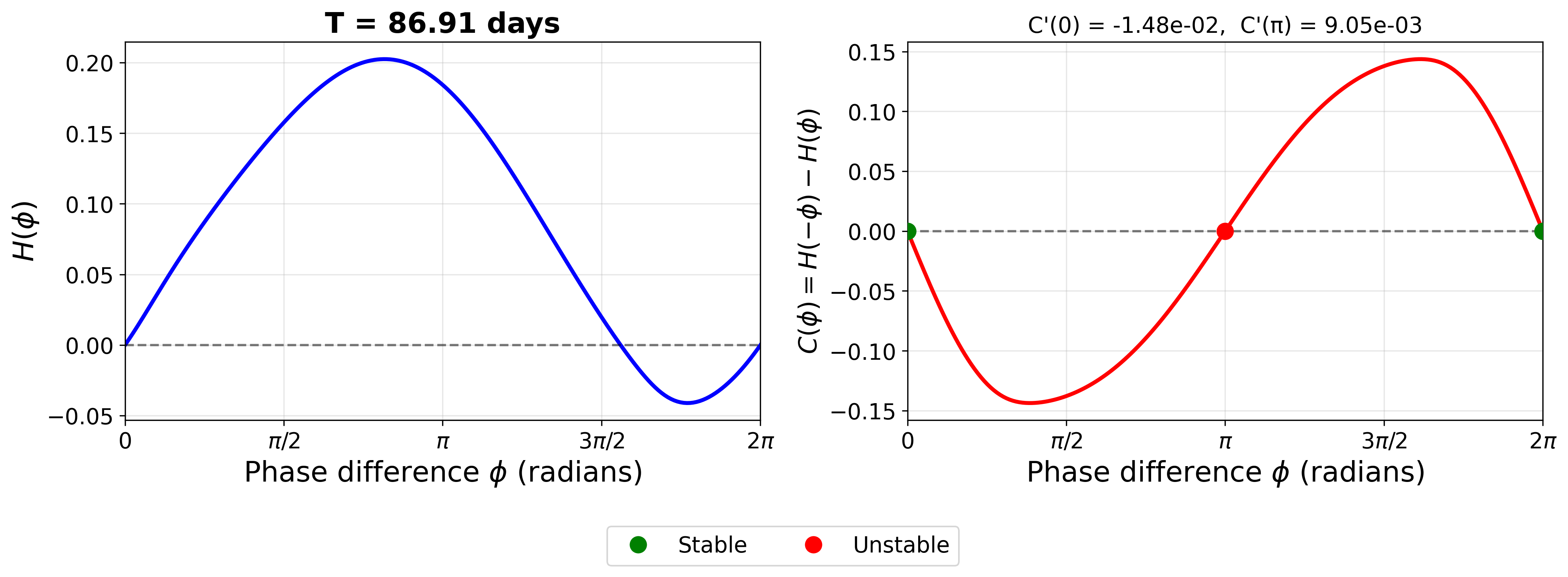}
    \caption{Coupling function $C(\phi)$ and fixed point stability for the Physical Only model. The model does not exhibits a bifurcation at $\phi = \pi$; anti-phase is unstable across all parameter values, and synchrony ($\phi = 0$) is the unique stable fixed point.}
    \label{fig:phase_reduction_physical_coupling}
\end{figure}

\begin{figure}[htbp]
    \centering
    \includegraphics[width=\textwidth]{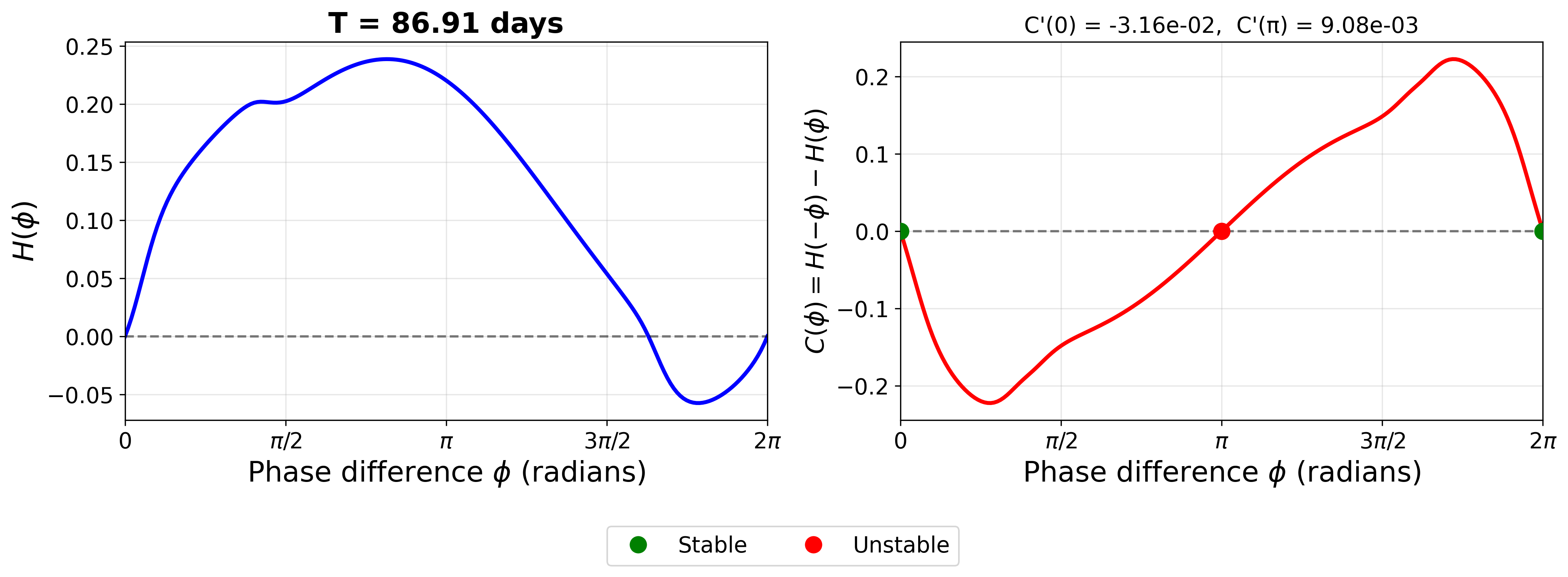}
    \caption{Coupling function $C(\phi)$ and fixed point stability for the Combined model. Like the Physical Only model, this model does not exhibits a bifurcation at $\phi = \pi$; anti-phase is unstable across all parameter values, and synchrony ($\phi = 0$) is the unique stable fixed point.}
    \label{fig:phase_reduction_combined_coupling}
\end{figure}

The phase reduction analysis for the Physical Only and Combined Coupling models follows the same procedure as the Social Only model. Note that the iPRC $Z(t)$ is identical across all three models since it is computed from the uncoupled oscillator, which is the same in each case. The coupling functions $G_j(X_j, X_k)$ for these models were derived in \eqref{G_j_for_other_two}, and the numerical computation of $H(\phi)$ and $C(\phi)$ proceeds analogously.
Figure \ref{fig:phase_reduction_physical_coupling} and Figure \ref{fig:phase_reduction_combined_coupling} show the numerical calculation of $H(\phi)$ and $C(\phi)$ for physical coupling model and combined coupling model respectively. \\
The absence of bifurcations in these models can be understood by examining the relative magnitudes of the iPRC components. We find that $|Z_I| \gg |Z_v|$ throughout the limit cycle. Since both the Physical Only and Combined Coupling models include coupling through the $I$ component—via the term $\beta S_j (I_k/\psi_k)$ in \eqref{G_j_for_other_two}—this coupling is amplified by the large $Z_I$ component. The resulting contribution to $H(\phi)$ strongly favors synchronization, overwhelming any tendency toward anti-phase coordination that might arise from the social coupling term. In contrast, the Social Only model couples exclusively through the $v$ component, which is modulated by the smaller $Z_v$, allowing for the richer bifurcation structure observed in Section~\ref{bifurcation_kappa}.

\section{Bifurcation Diagrams for Other Parameters}

\begin{figure}
    \centering
    \includegraphics[width=\textwidth]{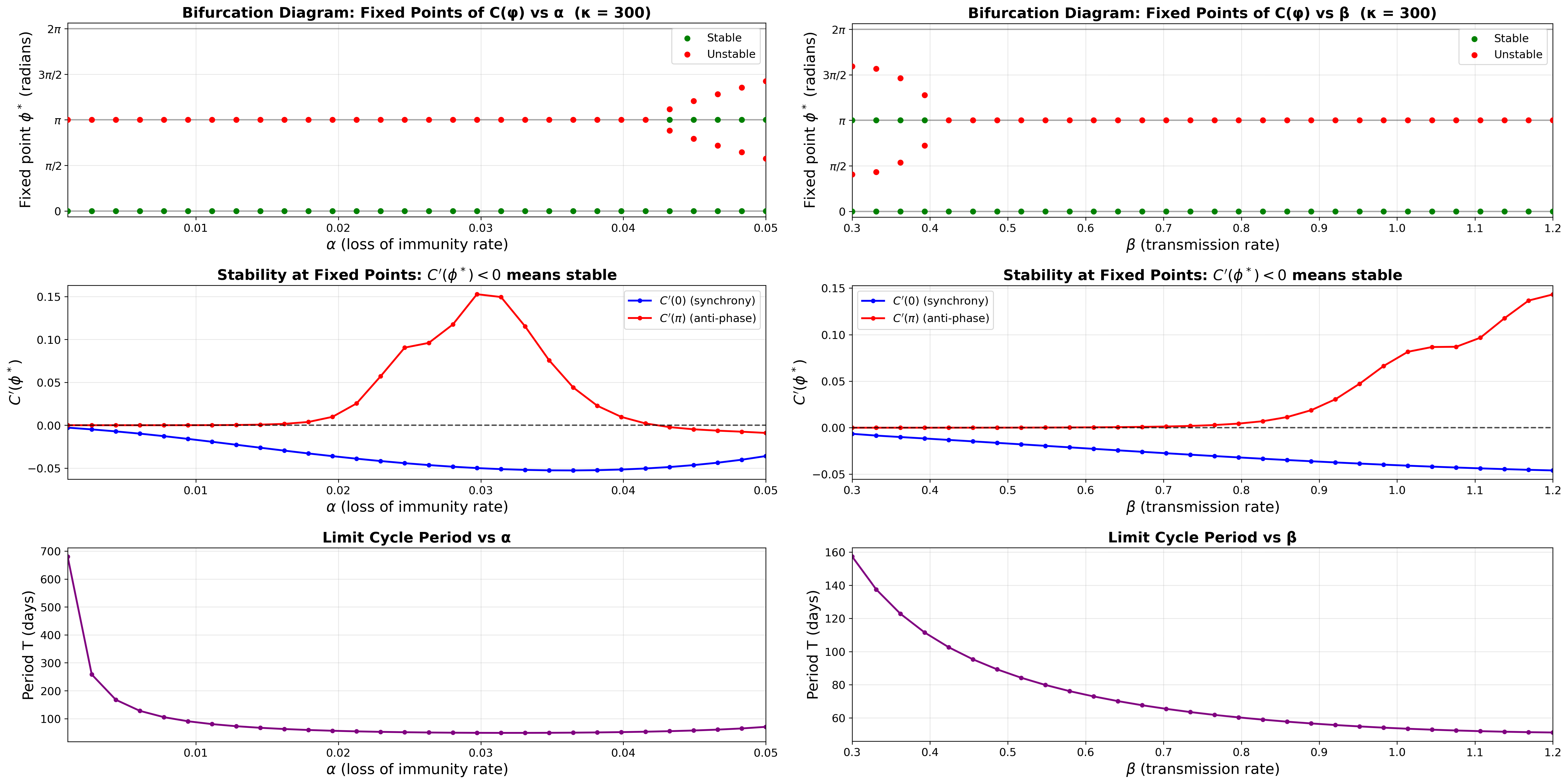}
    \caption{Bifurcation diagrams for parameters $\alpha$ and $\beta$ in the Social Only Coupling model. Note that $\beta$ shows the reverse direction compared to other parameters that exhibit subcritical pitchfork bifurcations}
    \label{fig:alpha_beta_bifurcation}
\end{figure}

\begin{figure}
    \centering
    \includegraphics[width=\textwidth]{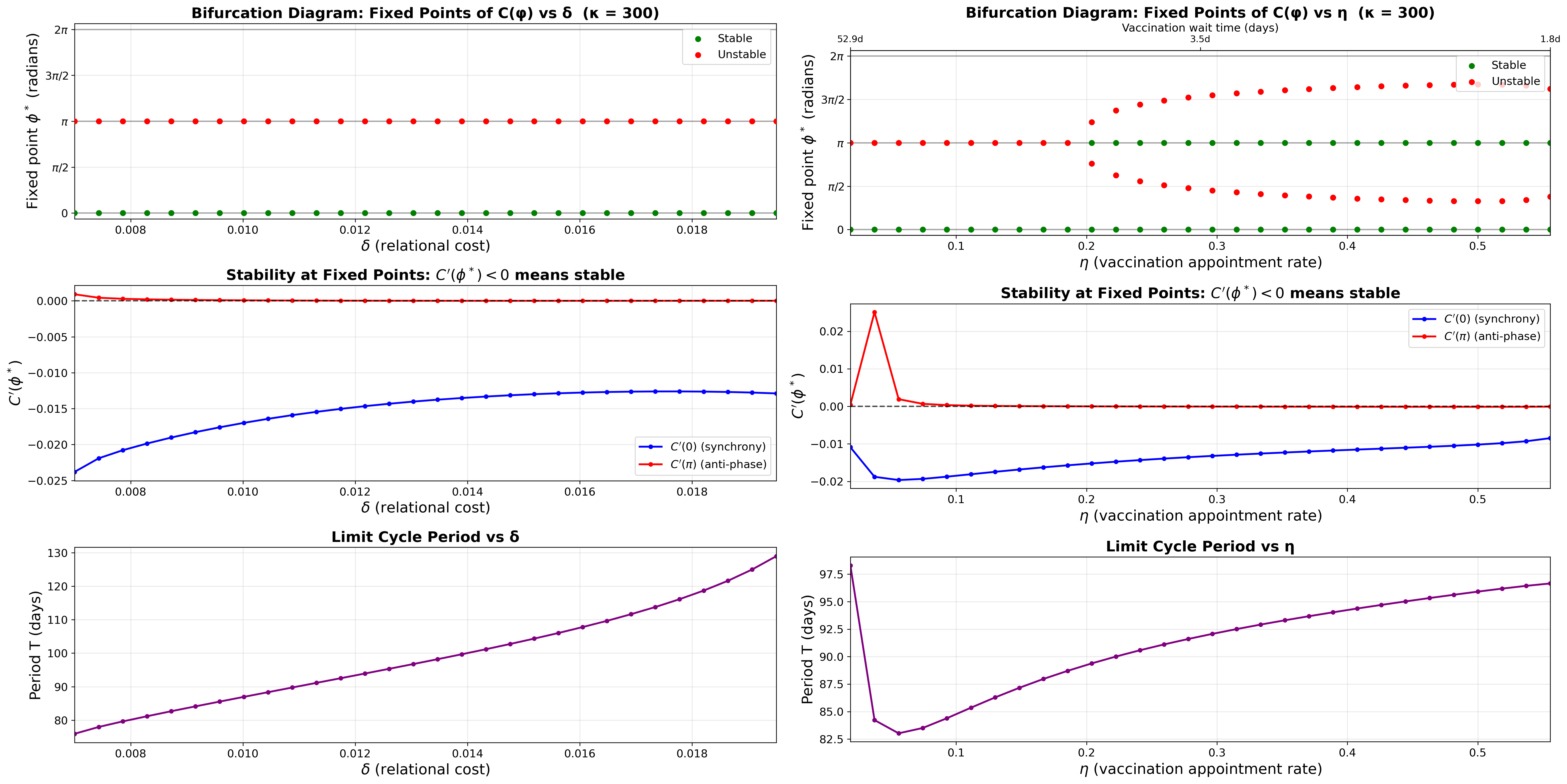}
    \caption{Bifurcation diagrams for parameters $\delta$ and $\eta$ in the Social Only Coupling model.}
    \label{fig:delta_eta_bifurcation}
\end{figure}

\begin{figure}
    \centering
    \includegraphics[width=\textwidth]{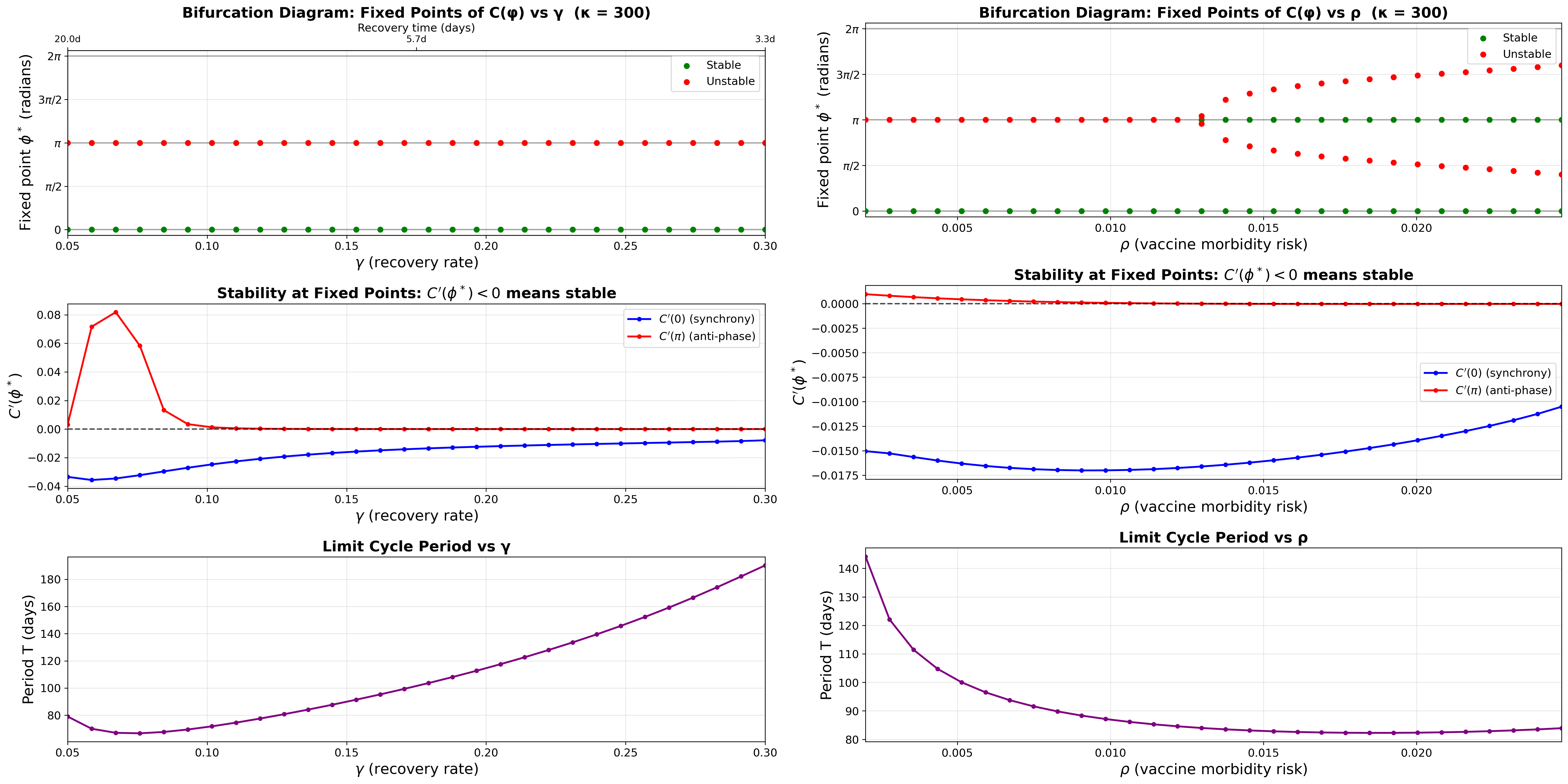}
    \caption{Bifurcation diagrams for parameters $\delta$ and $\eta$ in the Social Only Coupling model.}
    \label{fig:gamma_rho_bifurcation}
\end{figure}

Beyond $\kappa$, several other parameters in the Social Only Coupling model also induce bifurcations at the anti-phase fixed point $\phi = \pi$. Figure~\ref{fig:alpha_beta_bifurcation}, \ref{fig:delta_eta_bifurcation}, and \ref{fig:gamma_rho_bifurcation} shows the bifurcation diagrams for parameters $\alpha, \beta, \delta, \eta, \gamma$ and $\rho$. Each of these parameters exhibits a subcritical pitchfork bifurcation similar to that observed for $\kappa$, where anti-phase transitions from unstable to stable as the parameter crosses a critical threshold.

Notably, $\beta$ displays the reverse bifurcation direction: increasing $\beta$ destabilizes anti-phase, whereas for the other parameters ($\kappa$, $\rho$, $\alpha$, $\eta$), increasing the parameter stabilizes anti-phase. In contrast, parameters $\gamma$ and $\delta$ do not produce bifurcations; anti-phase remains unstable across their entire feasible range.

\end{document}